# Optically-detected NMR of optically-hyperpolarized $^{31}$P neutral donors in $^{28}$Si

M. Steger,[1] T. Sekiguchi,[1] A. Yang,[1] K. Saeedi,[1] M. E. Hayden[1], M. L. W. Thewalt,[1*]
K. M. Itoh,[2] H. Riemann,[3] N. V. Abrosimov,[3] P. Becker,[4] and H.-J. Pohl[5]

[1]*Department of Physics, Simon Fraser University, Burnaby, BC, Canada V5A 1S6*
[2]*School of Fundamental Science and Technology, Keio University, Yokohama 223-8522, Japan*
[3]*Leibniz-Institut für Kristallzüchtung, 12489 Berlin, Germany*
[4]*PTB Braunschweig, 38116 Braunschweig, Germany*
[5]*VITCON Projectconsult GmbH, 07743 Jena, Germany*

The electron and nuclear spins of the shallow donor $^{31}$P are promising qubit candidates invoked in many proposed Si-based quantum computing schemes. We have recently shown that the near-elimination of inhomogeneous broadening in highly isotopically enriched $^{28}$Si enables an optical readout of both the donor electron and nuclear spins by resolving the donor hyperfine splitting in the near-gap donor bound exciton transitions. We have also shown that pumping these same transitions can very quickly produce large electron and nuclear hyperpolarizations at low magnetic fields, where the equilibrium electron and nuclear polarizations are very small. Here we show preliminary results of the measurement of $^{31}$P neutral donor NMR parameters using this optical nuclear hyperpolarization mechanism for preparation of the $^{31}$P nuclear spin system, followed by optical readout of the resulting nuclear spin population after manipulation with NMR pulse sequences. This allows for the observation of single-shot NMR signals with very high signal-to-noise ratio under conditions where conventional NMR is not possible, due to the low concentration of $^{31}$P and the small equilibrium polarization.

Recent proposals [1-5] to use the electron and nuclear spins of shallow donor impurities as qubits for Si-based quantum computing (QC) have led to renewed interest in the properties of these systems [5-17]. Most of these studies focused on phosphorus, the most common shallow donor in Si, which has the added advantage of having only one stable isotope, $^{31}$P, with an $I = ½$ nuclear spin. Many of these QC schemes involve enriched $^{28}$Si as the host material, due to the elimination of the nonzero $^{29}$Si nuclear spin, and the resulting greatly improved coherence times. It has recently been shown [6] that the elimination of inhomogeneous isotope broadening inherent in highly enriched $^{28}$Si has the added benefit of enabling an optical measurement of the donor electron and nuclear spins using resolved hyperfine components in the donor bound exciton (D$^0$X) transitions [7], and furthermore allows for the rapid hyperpolarization of both electron and nuclear spins at low magnetic fields by resonant optical pumping [10].

Here we show preliminary results of NMR experiments performed on dilute $^{31}$P in $^{28}$Si using optical excitation of the D$^0$X hyperfine components to initially hyperpolarize the nuclear spin, and optical readout of the change in nuclear spin brought about by the application of RF magnetic fields at the NMR frequencies. Previous conventional NMR studies of $^{31}$P in Si were restricted to very heavily doped samples [18-20]. Our method provides information in a regime inaccessible to conventional NMR due to both the small number of spins and the low equilibrium polarization. We also obtain very high signal-to-noise ratios, allowing single-shot readout without the need for signal averaging. At higher magnetic fields, it is possible to detect neutral donor NMR indirectly in dilutely-doped samples using ENDOR methods [5, 14, 16], while at very low fields magnetic resonance has been detected by electric transport [17].

The spin Hamiltonian of an isolated $^{31}$P neutral donor (D$^0$) in a magnetic field $B$ along the $z$ axis can be written:

$$H_{Si:P} = g_e\mu_B BS_z - g_n\mu_n BI_z + a\mathbf{S}\cdot\mathbf{I}, \tag{1}$$

where $\mathbf{S}$ and $\mathbf{I}$ are the electron and $^{31}$P nuclear spins, respectively and $a$ is the hyperfine constant. In frequency units, $g_e\mu_B/h = \gamma_e = 2.79715(14)$ MHz/G, $g_n\mu_n/h = \gamma_n = 1.725145(46)$ kHz/G, and $a/h = A = 117.53(2)$ MHz [21-23]. It is also useful to define $\gamma_\pm = \gamma_e \pm \gamma_n$. At zero field the hyperfine interaction splits the donor ground state into a triplet and a singlet separated by $a$, while for $B > 0$ there are four resolved hyperfine states:

$$|1\rangle = |\uparrow\Uparrow\rangle \tag{2}$$

$$|2\rangle = \cos(\eta/2)|\uparrow\Downarrow\rangle + \sin(\eta/2)|\downarrow\Uparrow\rangle \tag{3}$$

$$|3\rangle = |\downarrow\Downarrow\rangle \tag{4}$$

$$|4\rangle = \cos(\eta/2)|\downarrow\Uparrow\rangle - \sin(\eta/2)|\uparrow\Downarrow\rangle \tag{5}$$

where $\uparrow(\downarrow)$ denotes electron spin +1/2 (–1/2), $\Uparrow(\Downarrow)$ denotes nuclear spin +1/2 (–1/2), and $\tan(\eta) = A/(\gamma_+ B)$. For simplicity, in the following we denote the mixed states $|2\rangle$ and $|4\rangle$ by the spin eigenstates to which they tend in the limit of high $B$, $|\uparrow\Downarrow\rangle$ and $|\downarrow\Uparrow\rangle$, respectively (at the field of interest here, $\eta$ is only ~2.8°). In frequency units, the energies of these states can be written in descending order as:

$$|\uparrow\Uparrow\rangle : (\gamma_- B + A/2)/2 \qquad (6)$$

$$|\uparrow\Downarrow\rangle : \left(\sqrt{(\gamma_+ B)^2 + A^2} - A/2\right)/2 \qquad (7)$$

$$|\downarrow\Downarrow\rangle : (-\gamma_- B + A/2)/2 \qquad (8)$$

$$|\downarrow\Uparrow\rangle : -\left(\sqrt{(\gamma_+ B)^2 + A^2} + A/2\right)/2 \qquad (9)$$

(the order of the $|\uparrow\Uparrow\rangle$ and $|\uparrow\Downarrow\rangle$ states is reversed above $B \sim 3.4$ T, but this is far above our field of interest). The NMR transitions studied here are between $|\uparrow\Uparrow\rangle$ and $|\uparrow\Downarrow\rangle$, and between $|\downarrow\Downarrow\rangle$ and $|\downarrow\Uparrow\rangle$, with frequencies $RF_\uparrow$ and $RF_\downarrow$, respectively. $RF_\uparrow$ will have a maximum value, $RF_{\uparrow max}$, and $RF_\downarrow$ a minimum value, $RF_{\downarrow min}$, at a field $B_0$:

$$B_0 = A \frac{\gamma_-}{\gamma_+ \sqrt{\gamma_+^2 - \gamma_-^2}} = 844.91(15) \text{ G.} \qquad (10)$$

This is an advantageous field for NMR measurements, since the effects of any inhomogeneity in the magnetic field are minimized, as was recognized earlier for the very analogous atomic hydrogen system [24]. At any given field, $RF_\uparrow + RF_\downarrow = A$, while at $B_0$, $\Delta RF = RF_\downarrow - RF_\uparrow$ has a minimum:

$$\Delta RF_{min} = A\sqrt{1 - \left(\frac{\gamma_-}{\gamma_+}\right)^2} = 5.834(1) \text{ MHz} \qquad (11)$$

or a predicted $RF_{\downarrow min} = 61.68(1)$ MHz and $RF_{\uparrow max} = 55.85(1)$ MHz. The change in the NMR frequencies for small deviations in $B$ near $B_0$ can also be calculated from these parameters and is $\pm 2.0380(4)$ HzG$^{-2}$.

The D$^0$X transitions are described in Fig. 1, showing the allowed transitions including hyperfine splittings numbered from 1 to 12 in order of increasing energy, as described previously [7], together with a diagram indicating the origin of the nuclear hyperpolarization observed [10] for the specific case of pumping line 6. Note that our photoluminescence excitation spectroscopy (PLE) measurements are essentially absorption measurements, connecting D$^0$ initial states and D$^0$X final states, so the relative intensities give directly the relative populations in the four D$^0$ hyperfine states when comparing transitions having the same selection rules, such as for example lines 5, 6, 7 and 8.

The optical systems used to obtain the PLE spectra and the resonant optical hyperpolarization have been described previously [7, 10]. The same $^{28}$Si sample is used in these studies, a disk ~10 mm in diameter and 1.5 mm thick, enriched to 99.991% and containing 1x10$^{14}$ cm$^{-3}$ boron and 7x10$^{14}$ cm$^{-3}$ phosphorus, with (001) faces perpendicular to **B**, and optical excitation and luminescence collection from the edge of the sample. The sample is loosely mounted in the center of a copper coil which can be tuned and matched to present a 50 ohm load at either RF frequency, and immersed in liquid He. The loaded Q of the resonant circuit was 115. The RF **B** field was perpendicular to the static field and parallel to the direction of laser excitation and photoluminescence detection. The temperatures given here are those of the He bath as inferred from the vapour pressure above the bath, which was not closely regulated. During the pulsed NMR measurements, the vapour pressure was either 1.0(1) Torr or 1 atm, corresponding to temperatures of 1.3 K and 4.2 K, respectively. The $B$ field was provided by an iron core electromagnet with 15.2 cm diameter pole faces separated by 9.5 cm, and monitored at the center of one pole face with a gaussmeter (Group3 DTM-151). The stability and reproducibility of $B$ were better than 0.1 G, but the field at the sample could only be estimated within ±0.5 G. The RF was provided by an Agilent E4420B signal generator. Part way through these studies the frequency accuracy was improved, to better than one part in 10$^{10}$, by providing a 10 MHz reference from a GPS-conditioned Rb clock. For all these studies the output of the synthesizer was used without amplification, and for pulsed NMR it was switched on and off using high isolation electronic switches driven by pulses of adjustable widths and delays generated with a simple programmable USB counter/timer module. With an RF power of +13 dBm (20 mW) delivered into the resonant circuit matched to 50 ohms, the π/2 pulse width was 25 µs.

Typical PLE spectra are shown in Fig. 2, which is very similar to Fig. 2 of our earlier work on optical hyperpolarization [10], except for the ~2 times higher $B$ field. At the bottom is the unpumped spectrum, showing the 12 D$^0$X hyperfine components, and the expected absence of significant equilibrium electron or nuclear polarization. Above are the two hyperpolarized spectra relevant to this study, showing the large electron and nuclear polarization obtained by pumping line 6, which hyperpolarizes the system into $|\downarrow\Downarrow\rangle$, or line 8, which hyperpolarizes the system into $|\uparrow\Uparrow\rangle$. Also shown are the probe transitions, selected to give the maximum possible increase in signal when the nuclear hyperpolarization is reduced by NMR, namely line 4 when pumping line 6, and line 10 when pumping line 8 (due to the very strong hyperpolarization of the electron spin [10], it is only possible to observe NMR for the electron spin opposite to the one being pumped). Unlike conventional NMR, which detects the precession of the net magnetization in the *x-y* plane, our optical method is sensitive only to $I_z$, the nuclear polarization parallel to ***B***.

Our investigation of optically polarized/optically detected NMR began with the simplest possible arrangement, which we refer to as CW NMR. Here the pump and probe lasers are set to pump 6/probe 4 (pump 8/probe 10) and left on continuously, while the RF signal driving the coil, which is also on continuously, is scanned across the expected $RF_\downarrow$ ($RF_\uparrow$) resonance frequency, and the PLE signal resulting from the probe laser is monitored. An example of a CW NMR resonance for the $RF_\uparrow$ branch is shown in Fig. 3. The resonance could be easily detected even when hand-scanning the RF frequency across the resonance, but note that very low RF power levels had to be used in order to avoid power broadening of the resonance, as indicated in Fig. 3. Similar scans were conducted for both branches and for several *B* fields around the expected optimal $B_0$ in order to test our field calibration, as summarized in Fig. 4. The observed values for $B_0$, $RF_{\uparrow\text{max}}$, $RF_{\downarrow\text{min}}$, $\Delta RF_{\text{min}}$ and the curvature at $B_0$ obtained by fitting this data were all in reasonable agreement with the predicted values. However, we emphasize that the values of $RF_{\uparrow\text{max}}$ and $RF_{\downarrow\text{min}}$ determined by CW NMR and shown in Fig. 4 are not as precise as those determined later by pulsed NMR, for several reasons. First, the Rb clock reference was added after the CW NMR measurements. Second, the requirement of very low RF power during the CW NMR measurements resulted in very long scan times, and spectra which were skewed depending on the direction of the RF scan. Finally, having both lasers on during the application of the RF will produce some sample heating, which we will later show may affect the precise value of *A*.

The scheme used for measuring pulsed NMR is outlined in Fig. 5. Both lasers are cycled on for 8 s and then off for 2 s. The 8 s on time is approximately twice the time needed to fully hyperpolarize the system at this field. In the absence of any RF, when the lasers are switched on again after the 2 s off time, the probe signal rises immediately to its previous value, since the nuclear $T_1$ in the absence of optical excitation is very long [10]. However, if resonant RF is applied during the dark interval, the nuclear hyperpolarization will be reduced, and when the lasers are switched on there will be a transient excess signal, which will then be repolarized to the original level. It is the area under this transient which is the optical NMR signal, and for each shot it is normalized against the signal after the transient in order to reduce the effects of drift. The magnetic field is set to $B_0$ as determined by the CW NMR measurements.

The simplest pulse scheme uses two $\pi/2$ pulses, with variable separation $\tau$, as shown in Fig. 5(a). In order to obtain a clearer signature, the RF frequency was offset by ±1 kHz from the $RF_{\uparrow\text{max}}$ or $RF_{\downarrow\text{min}}$ determined previously by CW NMR. This should result in fringes with a period of 1 ms as $\tau$ is scanned, and we refer to this experiment as a Ramsey fringe measurement due to the strong fringes as shown in Fig. 6, but in reality what is being measured is simply a free induction decay, together with the offset between the applied RF frequency and the actual resonance frequency. During these measurements we discovered that what was thought originally to be a drift in the measured frequencies was in fact a surprisingly significant change in *A* with temperature and pressure. These are coupled in our system, since we reduce the temperature below 4.2 K by reducing the vapour pressure of the liquid He, resulting in a combined effect, as originally realized in a study of shifts in the $^{28}$Si band gap energy below 4.2 K and 1 atm [25]. The size of this change in resonant frequency is readily apparent in Fig. 6. By fitting the data, accurate values of the fringe frequency and the amplitude decay time, $T_2^*$, are obtained. It should be emphasized that each data point in Fig. 6 comes from a single measurement, with no signal averaging.

The frequency data for our standard low temperature/low pressure conditions of 1.3 K and 1.0(1) Torr are summarized in Table I. Note that the fringe frequencies are not exactly 1 kHz, showing that the resonance frequencies determined by the CW NMR method described earlier were too low by ~25 (~27) Hz for $RF_\uparrow$ ($RF_\downarrow$), for reasons which have already been discussed. In any case, the fringe frequencies for each branch sum very precisely to 2 kHz, as they should, giving an indication of the accuracy of the frequency determination. Table I also shows $RF_{\uparrow\text{max}}$ and $RF_{\downarrow\text{min}}$ as determined by the two-pulse measurement, and their sum, *A*, the $^{31}$P hyperfine constant, consistent with the previous [21] value of 117.53(2) MHz, but with greatly improved accuracy. At $T$ = 4.2 K and 1 atm of pressure, *A* is reduced by 3116 Hz, or 26.5 ppm, an effect which would have been invisible in previous measurements. We find this change to be surprisingly large, although we have no explanation for its magnitude at this time.

It is also interesting that the results listed in Table I give a value for $\Delta RF_{\text{min}}$ = 5,830,462.3(10) Hz, which is significantly different from the 5.834(1) MHz predicted in Eq. (11), a discrepancy which is barely affected by substituting our improved value of *A*. A measurement of $\Delta RF_{\text{min}}$ is in essence a precise determination of $\gamma_n$ relative to $\gamma_e$, giving $\gamma_n/\gamma_e$ = 6.1606748(28) × 10$^{-4}$. While $\gamma_e$ was determined using the $g_e$ appropriate for $^{31}$P in Si [21], $\gamma_n$ was calculated using the $g_n$ of the free $^{31}$P nucleus [22]. Thus our measurements of *A* and $\Delta RF_{\text{min}}$ can be used to obtain a value of $\gamma_n$ specific for the $^{31}$P D$^0$ in Si. Combining the previously measured value of $\gamma_e$ [21, 23] with our observed values for *A* and $\Delta RF_{\text{min}}$ gives a value of $\gamma_n$ = 1.72323(9) kHz/G for the $^{31}$P D$^0$ in Si, or a chemical shift of −1110(55) ppm relative to the free nucleus.

While a precise value for *B* was not required for the above calculation, it would be desirable for other precision measurements. Note that our new values for *A* and $\gamma_n$ can be used in Eq. (10) to calculate an improved value of $B_0$ =

845.34 G, which agrees with the original prediction to within the limited accuracy of our estimation of $B$ at the sample location. This field can in principle be determined from the separations of lines 1 and 7, or 5 and 11, in the PLE spectrum, which both equal $\gamma_e B$. This gives a value of $B_0 =$ 845.2 G, closer to the new value than the value given in Eq. (10), but at this time we have no independent means of determining the accuracy of the PLE energy scale, which is set by our tuneable laser locking and scanning system. A very desirable future improvement would be a second RF coil to enable the simultaneous measurement of the NMR and EPR frequencies.

The amplitude decay fits for the four two-pulse scans at $T = 1.3$ K listed in Table I gave results consistent with $T_2^* = 16(1)$ ms. At this time we do not have sufficient data at $T = 4.2$ K to know whether the longer $T_2^*$ seen at the top of Fig. 6 is a real effect or instead reflects the greatly increased noise in the 4.2 K spectra caused by the boiling liquid He bath. The $T_2^*$ results include contributions from all possible dephasing mechanisms, including $B$ field inhomogeneities and the inequivalence of different $^{31}$P sites. In order to remove the effects of static dephasing and measure $T_2$, we implemented a simple three-pulse Hahn echo measurement, as shown in Fig. 5(b). For this experiment the RF frequency was set exactly to $RF_{\downarrow\text{min}}$ or $RF_{\uparrow\text{max}}$ as determined by the previous two-pulse measurements, and $B$ was set to $B_0$. To avoid the drift problems inherent in DC measurements, for each delay two measurements were taken, one with all three pulses having the same RF phase, and one in which the phase of the first $\pi/2$ pulse was inverted, with the reported signal being the difference between these two measurements.

The results of the Hahn echo measurement on the $RF_{\downarrow}$ branch are shown in Fig. 7. A measurement on the $RF_{\uparrow}$ branch gave identical results within the noise level. The $T_2$ coherence time was found to be ~230 ms for both branches. At present we have no explanation for the apparently slower decay at short times. While 230 ms is a long coherence time by most standards, it is significantly shorter than the 1.75 s reported for $^{31}$P in $^{28}$Si as measured by an ENDOR technique at a higher field [5]. While we are only beginning to explore the many experimental parameters which may affect the value of $T_2$ measured using our method, it seems likely that a dominant one may be the sample itself, especially the $^{28}$Si enrichment and the $^{31}$P concentration, the latter being considerably larger in our sample than the one studied by ENDOR [5].

In conclusion, we have reported preliminary measurements of NMR data from highly enriched $^{28}$Si lightly doped with $^{31}$P using optical hyperpolarization of the spin system before application of the RF, and optical measurement of the change in nuclear spin caused by the RF. We emphasize that these are very preliminary results, and should not be taken as the ultimate limits of the technique, or of parameters such as nuclear coherence times. Nevertheless, we have already succeeded in measuring the $^{31}$P hyperfine constant to more than 10,000 times the previous accuracy, and as a result have detected a significant shift with temperature and pressure which as yet has not been explained. A chemical shift of the $^{31}$P NMR resonance frequency in $^{28}$Si has been determined, and additional precision measurements are possible in the future. We have also measured a nuclear $T_2$ of 230 ms, and future work will focus on determining what limits this value, and to what extent it can be extended.

**Acknowledgements:** The authors thank John Morton, Thaddeus Ladd, Martin Brandt and Steve Lyon for many fruitful discussions. This work was supported by the Natural Sciences and Engineering Research Council of Canada (NSERC).

\* - thewalt@sfu.ca

**References**

[1] B. E. Kane, Nature **393**, 133 (1998).
[2] D. P. DiVincenzo, Fortschr. Phys. **48**, 771 (2001).
[3] R. Vrijen, E. Yablonovitch, K. Wang, H. W. Jiang, A. Balandin, V. Roychowdhury, T. Mor, and D. DiVincenzo, Phys. Rev. A **62**, 012306 (2000).
[4] A. M. Stoneham, A. J. Fisher, and P. T. Greenland, J. Phys. Condens. Matter **15**, L447 (2003).
[5] J.J.L. Morton, A.M. Tyryshkin, R.M. Brown, S. Shankar, B.W. Lovett, A. Ardavan, T. Schenkel, E.E. Haller, J.W. Ager, and S.A. Lyon, Nature **455**, 1085 (2008).
[6] D. Karaiskaj, M. L. W. Thewalt, T. Ruf, M. Cardona, H.-J. Pohl, G. G. Devyatich, P. Sennikov and H. Riemann, Phys. Rev. Lett. **86**, 6010 (2001).
[7] A. Yang, M. Steger, D. Karaiskaj, M. L. W. Thewalt, M. Cardona, K. M. Itoh, H. Riemann, N. V. Abrosimov, M. F. Churbanov, A. V. Gusev, A. D. Bulanov, A. K. Kaliteevskii, O. N. Godisov, P. Becker, H.-J. Pohl, J. W. Ager III, and E. E. Haller, Phys. Rev. Lett. **97**, 227401 (2006).
[8] N. Q. Vinh, P. T. Greenland, K. Litvinenko, B. Redlich, A. F. G. van der Meer, S. A. Lynch, M. Warner, A. M. Stoneham, G. Aeppli, D. J. Paul, C. R. Pidgeon and B. N. Murdin, PNAS **105**, 10649 (2008).
[9] D.R. McCamey, J. van Tol, G.W. Morley, and C. Boehme, Phys. Rev. Lett. **102**, 027601 (2009).
[10] A. Yang, M. Steger, T. Sekiguchi, M. L. W. Thewalt, T. D. Ladd, K. M. Itoh, H. Riemann, N. V. Abrosimov, P. Becker, and H.-J. Pohl, Phys. Rev. Lett. **102**, 257401 (2009).

[11] A. Yang, M. Steger, T. Sekiguchi,1 M.L.W. Thewalt, J.W. Ager III, and E.E. Haller, Appl. Phys. Lett. **95**, 122113 (2009).

[12] P. T. Greenland, S. A. Lynch, A. F. G. Van der Meer, B. N. Murdin, C. R. Pidgeon, B. Redlich, N. Q. Vinh and G. Aeppli, Nature **465**, 1057 (2010).

[13] T. Sekiguchi, M. Steger, K. Saeedi, M. L. W. Thewalt, H. Riemann, N. V. Abrosimov, and N. Nötzel, Phys. Rev. Lett. **104**, 137402 (2010).

[14] R. E. George, W. Witzel, H. Riemann, N. V. Abrosimov, N. Nötzel, M. L. W. Thewalt and J. J. L. Morton, Phys. Rev. Lett. **105**, 067601 (2010).

[15] M. H. Mohammady, G. W. Morley and T. S. Monteiro, Phys. Rev. Lett. **105**, 067602 (2010).

[16] G. W. Morley, M. Warner, A. M. Stoneham, P. T. Greenland, J. v. Tol, C. W. M. Kay and G. Aeppli, Nat. Mat. **9**, 725 (2010).

[17] H. Morishita, L. S. Vlasenko, H. Tanaka, K. Semba, K. Sawano, Y. Shiraki, M. Eto, and K. M. Itoh, Phys. Rev. B **80**, 205206 (2009).

[18] G. P. Carver, D. F. Holcomb and J. A. Kaeck, Phys. Rev. B **3**, 4285 (1971).

[19] H. Alloul and P. Dellouve, Phys. Rev. Lett. **59**, 578 (1987).

[20] M. Jeong, M. Song, T. Ueno, T. Mizusaki, A. Matsubara and S. Lee, J. Low Temp. Phys. **158**, 659 (2010).

[21] G. Feher, Phys. Rev. **114**, 1219 (1959).

[22] N. J. Stone, At. Data Nucl. Data Tables **90**, 75 (2005).

[23] P. J. Mohr and B. N. Taylor, Rev. Mod. Phys. **72**, 351 (2000).

[24] W. N. Hardy, A. J. Berlinsky and L. A. Whitehead, Phys. Rev. Lett. **42**, 1042 (1979).

[25] M. Cardona, T. A. Meyer and M. L. W. Thewalt, Phys. Rev. Lett. **92**, 196403 (2004).

| Branch | Applied RF freq. (Hz) | Fringe freq. (Hz) | Resonance freq. (Hz) |
|---|---|---|---|
| $RF_\uparrow$ | 55,845,712.00 | 1,024.9 | $RF_{\uparrow max}$ = 55,846,736.8(5) |
|  | 55,847,712.00 | 975.4 ($\Sigma$ = 2,000.3) |  |
| $RF_\downarrow$ | 61,676,172.00 | 1,027.2 | $RF_{\downarrow min}$ = 61,677,199.1(5) |
|  | 61,678,172.00 | 972.9 ($\Sigma$ = 2,000.1) |  |
|  |  |  | $A$ = 117,523,935.9(10) |
|  |  |  | $\Delta RF_{min}$ = 5,830,462.3(10) |

Table I: Results of the two-pulse 'Ramsey fringe' measurement at T=1.3 K and a He vapour pressure of 1.0(1) Torr, and the resulting values of $A$ and $\Delta RF_{min}$.

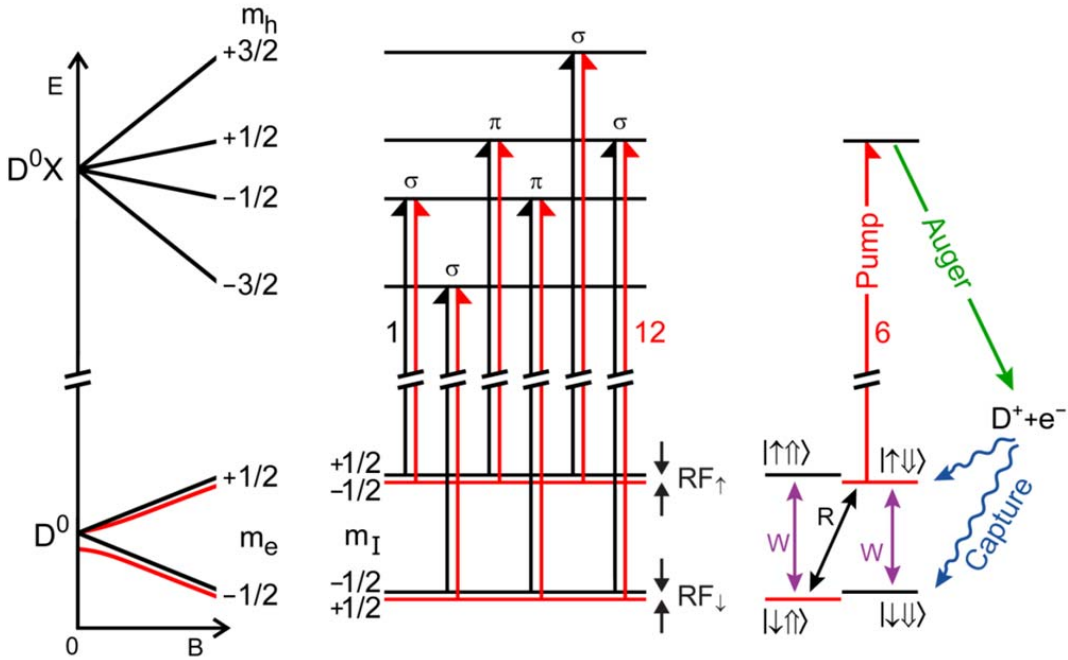

Figure 1: The Zeeman splitting for **B** ∥ [001] of the phosphorus neutral donor ($D^0$) and donor bound exciton ($D^0X$) ground states are shown on the left, in the low field regime where the nuclear Zeeman energy is small. The projection of the electron spin along the field direction is labelled by $m_e$, while $m_I$ labels the nuclear spin projection and $m_h$ the hole spin. The six dipole allowed bound exciton transitions, each of which is split into a doublet by the hyperfine interaction, are shown in the center, with components labelled from 1 to 12 in order of increasing energy, together with the NMR transitions, $RF_\uparrow$ and $RF_\downarrow$. On the right is a schematic of the resonant polarization process, showing how pumping transition 6 leads to a hyperpolarization of the donors into the $|\downarrow\Downarrow\rangle$ state [10].

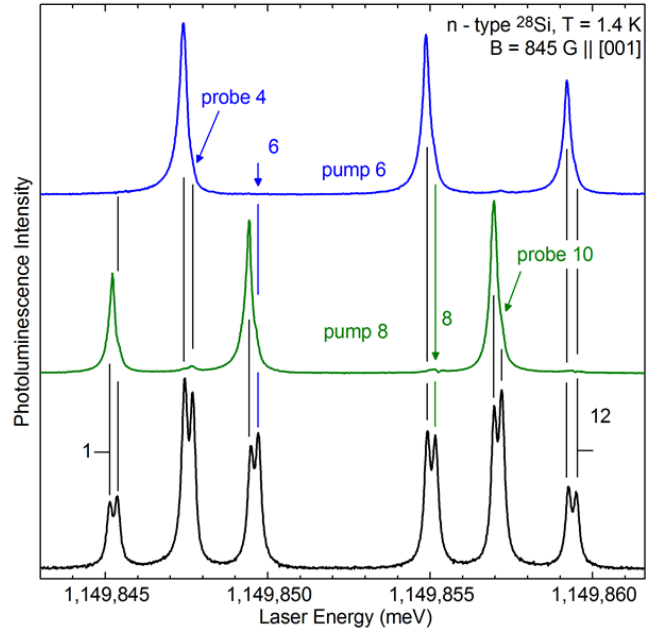

Fig. 2: The photoluminescence excitation (PLE) spectrum of the $^{31}$P-doped $^{28}$Si sample used in these studies is shown at the bottom without additional resonant pumping. Above are shown the PLE spectra with a strong pump laser on either transition 8 or 6, which give the largest hyperpolarizations of the donor electron and nuclear spins [10]. Also indicated are the two transitions which are used to monitor the redistribution of nuclear spins when measuring optically detected NMR: line 10 when pumping 8, and line 4 when pumping 6.

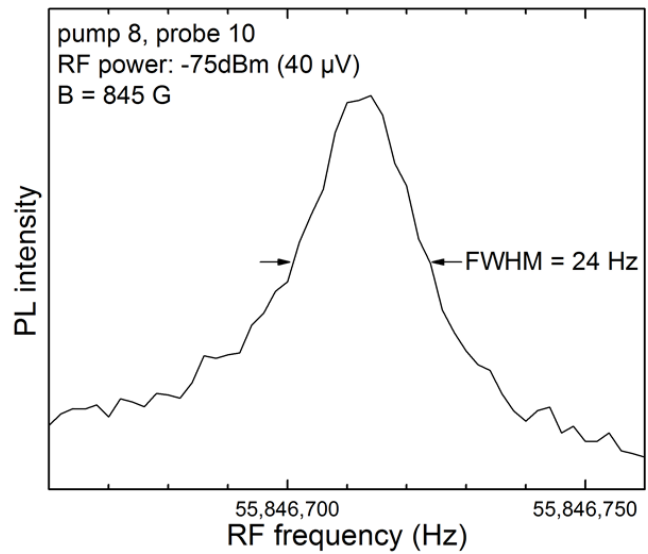

Fig. 3: Result of CW optically detected NMR showing an increase in the signal when probing line 10 and pumping line 8 as the RF frequency is tuned across the $RF_\uparrow$ resonance.

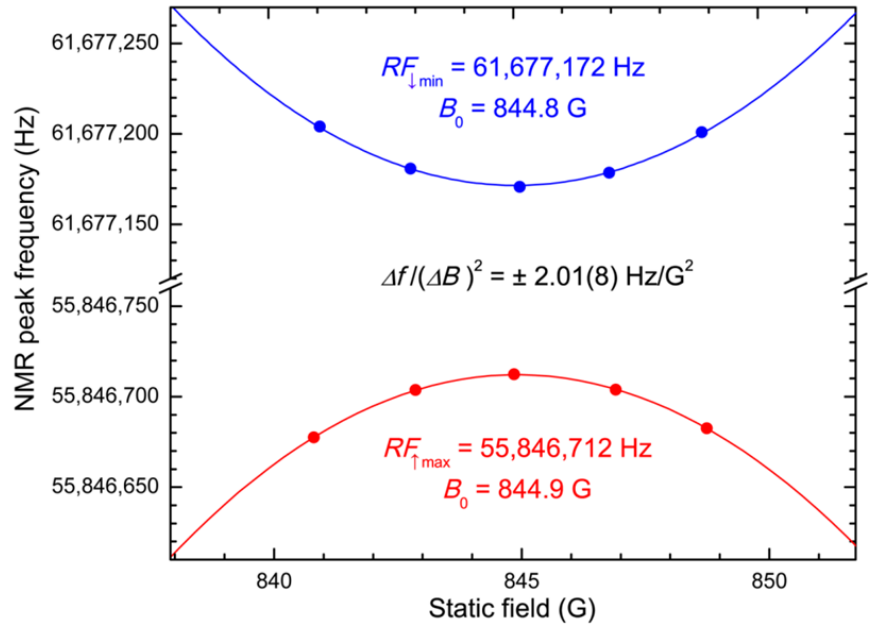

Fig. 4: The tuning curves for the two NMR resonance frequencies as a function of the applied magnetic field, determined by CW optically detected NMR. $RF_{\uparrow max}$, $RF_{\downarrow min}$, the corresponding magnetic field $B_0$, and the quadratic frequency shift relative to the extrema are obtained by fitting the data for each branch, as shown by the two solid lines.

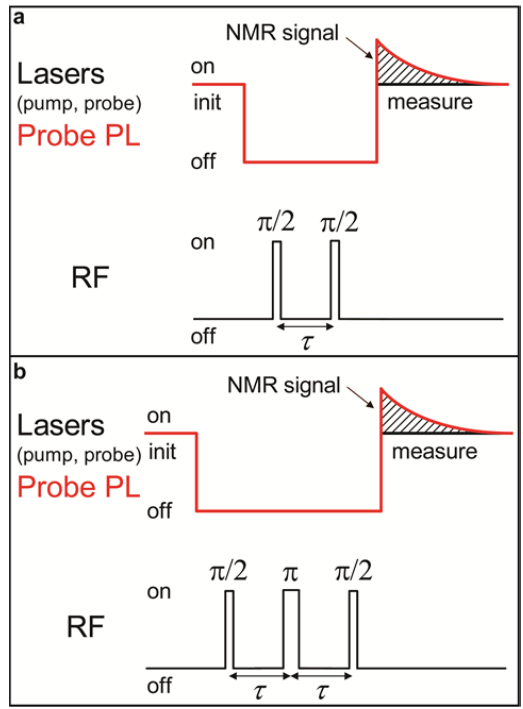

Fig. 5: The pulsed optically detected NMR scheme used here, at top for the two-pulse 'Ramsey fringe' experiment, and at bottom for the three-pulse Hahn echo measurement.

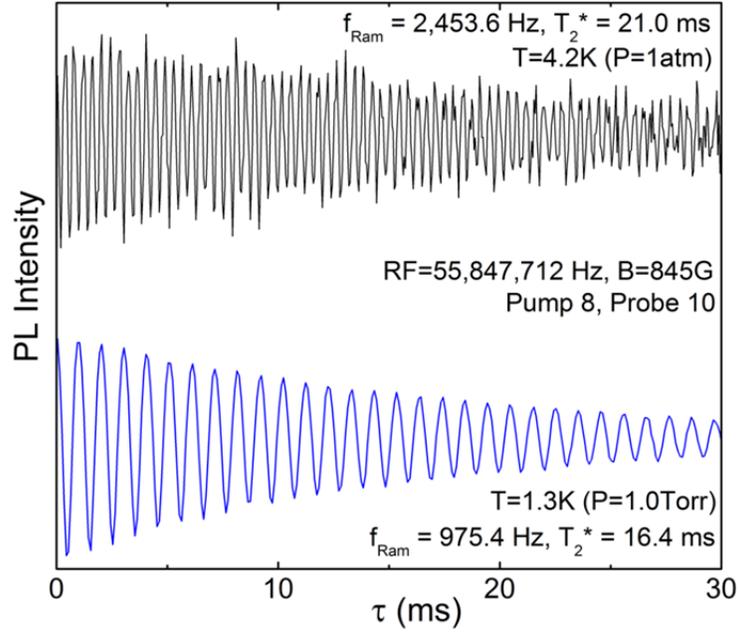

Fig. 6: The results for the two-pulse Ramsey fringe experiment on the electron spin up branch are shown both at our usual low temperature and pressure conditions as well as at a temperature of 4.2 K and atmospheric pressure. The results of fits to the fringe frequency, $f_{Ram}$, and amplitude decay time, $T_2^*$, are shown. The excess noise in the 4.2 K spectrum is due to the boiling He.

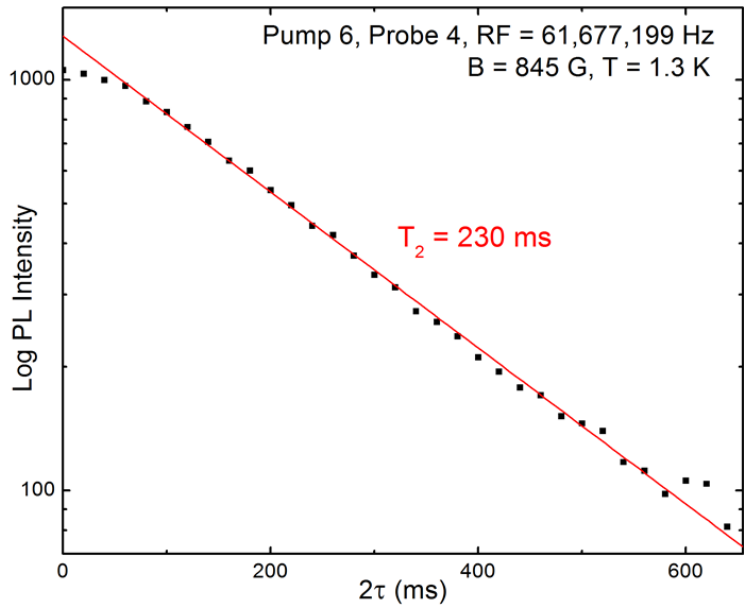

Fig. 7: The results of a three pulse Hahn echo measurement of the nuclear $T_2$ on the electron spin down branch at low temperature and pressure.